\documentclass[secnumarabic,amssymb, nobibnotes, aps, prd]{revtex4}%
\usepackage{graphicx}
\usepackage{dcolumn}
\usepackage{bm}
\usepackage{amsmath}%
\usepackage{amsfonts}%
\usepackage{amssymb}

\begin{document}

\title{Is there a galactic component for the ultra high energy cosmic rays?}

\author{C. E. Navia}
\address{Instituto de F\'{\i}sica, Universidade Federal Fluminense, 
Av. General Milton Tavares de Souza s/n, 
Gragoat\'{a} 24210-340, Niter\'{o}i, RJ, Brazil} 

\author{C. R. A. Augusto}
\address{Instituto de F\'{\i}sica, Universidade Federal Fluminense, 
Av. General Milton Tavares de Souza s/n, 
Gragoat\'{a} 24210-340, Niter\'{o}i, RJ, Brazil} 

\author{K. H. Tsui}
\address{Instituto de F\'{\i}sica, Universidade Federal Fluminense, 
Av. General Milton Tavares de Souza s/n, 
Gragoat\'{a} 24210-340, Niter\'{o}i, RJ, Brazil} 

\date{\today}

\begin{abstract}
Under the hypothesis that Gamma Ray Burst (GRB) might be responsible 
for the origin of Ultra High Energy Cosmic Ray (UHECR), we propose a 
two component (galactic and extra-galactic) model for the UHECR origin. 
The model is based on two facts. The first is the anisotropies found 
in the angular distribution of GRBs from BATSE catalog. Second is that, 
of all the located long-GRBs, only approximately 15 percent of them 
have their spectroscopic redshift determined, and some 38 percent of 
them have a x-ray, optical, or radio afterglow. So far, in short-GRBs, 
no afterglow and no red shift have been detected, suggesting that these 
GRB sources are inside or close to our Galaxy. This two component model 
for the UHECR is further supported by the experimental evidences of an 
UHECR excess around $10^{18}$ eV from the direction of the galactic 
central region. The model offers in a natural way an explanation for the 
presence of cosmic rays with energies beyond the Greisen-Zatsepin-Kuz'min 
(GZK) cutoff.

\end{abstract}

\pacs{PACS number: 96.40.De, 12.38.Mh,13.85.Tp,25.75.+r}

\maketitle

\section{Introduction}

After the discovery of the Microwave Background Radiation (MBR), that 
fills the universe as a sea of photons, with a mean temperature of 
2.7 K at present epoch, Greisen \cite{greisen66} and Zatsepin and 
Kuz'min \cite{zapsepin66} independently pointed out that the MBR would 
make the universe opaque to UHECR particles with energies approximately 
$5 \times 10^{19}$ eV and above. This hypothesis is well accepted and 
is known as the GZK cutoff. So far, observational data of several 
experiments about UHECR have report events with energies beyond the GZK 
cutoff \cite{hashida94,bird93,lawrence71}. In addition, the present data 
show strong discrepancies in the UHECR energy spectrum. On one side, 
the AGASA (extensive air shower) experiment claims a spectrum beyond the 
GZK cutoff, without any evidence of the GZK cutoff, on the other side the 
HIRES (air fluorescence) experiment is in agreement with the expected 
universal distribution of sources and showing the GZK cutoff in the energy 
spectrum. This means at least that there is a systematic error, probably 
in the energy determination, in one of the experiments. However, this 
discrepancy can also be attributed to latitude effects, due to the limited 
coverage of the sky by experiments located in the northern hemisphere. 
So far only experiments with a full (SUGAR experiment \cite{clay01}) or 
partial (AGASA experiment \cite{hashida99}) coverage of the galactic 
central region have reported a cosmic ray excess in the energy region of 
approximately $10^{18}$ eV in the direction of the galactic center. 
The accumulating data from AUGER (extensive air shower plus air fluorescence) 
experiment \cite{cronin00}, whose southern part is now in progress, on the 
basis of a larger data set will tell us which is the correct alternative 
in the near future.

In the past years, there are some surveys about a statistical correlation 
between UHECR events and compact sources at high redshift. Depending on 
the data set, in some cases a positive correlation has been reported 
\cite{dubovsky00}, but there are also negative results \cite{torres03}. 
The identification of compact objects like BL Lac or quasars at high 
redshift as UHECR sources reinforces a cosmological origin, and has taken 
to formulate new scenarios such as the violation of the Lorentz invariance 
\cite{coleman99} as responsible for the propagation of ultra high energy 
cosmic ray with energies above the GZK cutoff. In addition to the top-down 
models \cite{hill83}, where the collapse or decay of super-massive 
particles like magnetic monopoles, superconducting strings, as well as 
the $\nu$ Z burts \cite{fargion99} inside a volume with a radius less than 
50 MPc from the Earth can explain the UHECR data above the GZK cutoff. 
 
The hypothesis that Gamma Ray Burst (GRB) might be responsible for the 
origin of UHECR has been suggest earlier \cite{waxman95}. The GRBs are 
probably the most powerfully events in the universe, and it is believed 
that protons can be accelerate in a GRB by internal shocks taking place 
in a collimated jet direction. The GRBs have been observed in spacecraft 
experiments as short flashes of gamma rays that outshine the rest of the 
entire gamma-ray sky at a rate of around one event per day \cite{paciesas99}, 
mostly in the energy band of KeV to MeV . The first measurements of the 
redshifts in GRB afterglows \cite{costa97}, together with the highly 
isotropic distribution of their arrival directions, have established a 
cosmological origin for them. The association of GRBs and UHECRs is further 
supported by the similarity of energy generation rates. the average rate 
of gamma-ray energy emitted by GRBs is comparable to the energy generation 
rate of UHECRs. In addition, the almost isotropic distribution of the 
arrival directions of UHECR and GRBs is also an ingredient in favor of 
an association between them.

The main constraint on the association between GRB and UHECR is that 
UHECR events above the GZK cutoff require only GRBs inside a volume 
of radius less than 50 Mpc from the Earth can contribute to the UHECR 
flux. What is the expected rate of GRBs in this volume? A pessimistic 
answer on the basis of only cosmological origin to the GRBs gives an 
estimative of about one GRB per 100 years. On the other hand, from an 
optimistic point of view, statistical surveys have shown that it is 
not always possible to spectroscopically determine the red shift even 
in some long well located GRBs accompanied by x-ray, optical and radio 
afterglow. A plausible explanation for these results is the assumption 
of a local origin. That is to say, they are close to our Galaxy or 
inside it. Around 70 percent of the observed GRBs are the long-soft 
type with approximately 30 s duration. Some of these bursts have their 
measured red shifts clustering in $z\sim 1$, consequently they have a 
cosmological origin and they are connected with supernovae events. 
There is also another category of GRBs, the short-hard type with 
approximately 0.2 s duration, where no red shift and no afterglow have 
been detected from these bursts. Probably these short bursts result from 
a different engine than long bursts, they could be connected with compact 
binary mergers, like two neutron stars or with a black hole component, 
as well as binary pulsar. They are expected at a high rate of 
$2\times 10^{-4}$ per year in a galaxy like the Milk Way \cite{kalogera03}. 
These characteristics suggest the Two Component Model (TCM), galactic and 
extra-galactic for the origin of the GRBs.   

Another characteristic that strongly support the TCM for the GRBs origin 
is their isotropic distribution of the arrival direction. They are 
regarded to have a uniform distribution in galactic coordinates. 
A more accurate analysis of the angular distribution of GRB shows some 
deviations from a simple isotropic distribution \cite{manchanda05}. 
Under certain assumptions such as an Euclidean space and standard candle 
GRBs, the formulation of the TCM is plausible. Here we make an extension 
of these assumptions to the UHECR events. We point out here that a TCM 
for the UHECR is further supported by the experimental evidences of a 
UHECR particle excess around $10^{18}$ eV energies from the direction 
of the galactic central region. 

\section{The Seeliger's theorem and GRBs}

A consequence of the Seeliger's theorem is the following. If sources 
are uniformly distributed in a spherical volume $V$ of radius $r$, 
and have fixed brightness, then the flux $S(E)$ obtained from a source 
is proportional to $r^{-2}$, while the number $N$ of sources observed 
down to a given flux limit is proportional to $r^{3}$ 
\begin{equation}
N \propto S(E)^{-3/2}
\end{equation}
This relation has been observed in radioastronomy as the 
$\log N \rightarrow \log S(E)$ plot
\begin{equation}
\log N \propto -\beta \log S(E)
\end{equation}
where $\beta =3/2$ provided that the sources are homogeneously 
distributed in an Euclidian space. Then, if a true deviation from 
this value eventually takes place, we will have the evidence that 
the space is non-Euclidian or/and the sources are inhomogeneously 
distributed.

The black line in Fig.1 shows the $\log N \rightarrow \log S(E)$ plot 
obtained for the GRBs on the basis of 2704 burst from BATSE catalog. 
We can see from this figure that a simple $\beta=3/2$ (blue line) 
does not fit the data. This means that the spatial distribution of 
GRBs is not consistent with a homogeneous case. As already has been 
commented, a possible interpretation for this result is to invoke a 
non-Euclidian space. However, another alternative is to introduce a 
second population of GRBs with a gaussian distribution that reflects 
the radial distribution of matter in the galaxy and its surrounding 
halo. The fluencies of the observed GRBs between 
$10^{-5}$ to $10^{-7}$ $erg/cm^2$ imply isotropic burst energies up 
to approximately $4 \times 10^{54}$ erg. While taking beaming angle 
corrections of about 5 degrees into account, it is expected a narrow 
distribution around $5 \times 10^{50}$ erg. Here, we assume the same 
value as in \cite{manchanda05} that gives $10^{51}$ erg to the gaussian 
peak which corresponds to a mean flux of 2.8. The best fit is obtained 
for an amplitude of about 7.5 percent of the first component and a 
r.m.s. value of 13.4. In short, the TCM spatial burst distribution is 
given by the following cumulutive function
\begin{equation}
N(>S)=950 S^{-1.5}+70 e^{[-(S-2.8)/19]^2}.
\end{equation}\textsl{}
The red line in Fig.xx represent this function. 

\section{The extension to the UHECR}

Under the assumption of an association between GRBs and UHECR, the 
hypothesis of a second population of GRBs of galactic origin implies 
also the existence of UHECR of galactic origin. Due to the small 
statistics, especially in the energy region above the GZK cutoff, 
it is still not possible to mount the log N $\rightarrow$ log S(E) 
plot, only on the basis of UHECR data. Consequently, we make a 
calibration to obtain the TCM UHECR distribution on the basis of the 
TCM GRBs distribution. 

The detection of an excess of cosmic ray from the direction of the 
galactic center region observed by AGASA and SUGAR in the energy 
range of $10^{18}$ eV opens the door to the possibility that UHECR 
above the GZK might be created also in our Galaxy. We assume that 
the excess of cosmic ray from the direction of the galactic center 
in the energy range of $10^{18}$ eV corresponds to the peak of a 
gaussian distribution and represent the second population of the UHECR. 
This assumption permits us to make a calibration between the GRB flux 
and the energy of the UHECR. The other parameters of this second 
population distribution of UHECR are the same as the GRBs distribution. 
Figure 2 summarizes the situation where the cumulative distributions 
for UHECR and for three different energies in the peak of the gaussian 
distribution are shown. It is possible to see that the contribution of 
the second population to the UHECR flux around the GZK cutoff is 
sensitive to the value of the gaussian peak. The comparison of the 
predictions of the TCM and the experimental data is shown in Fig.3. 

\section{Conclusions}

Under the assumption that GRBs might be responsible for the origin of 
UHECR, we have presented here an extension of the Two Component Model 
of GRBs (galactic, and extra-galactic) to explain the origin of UHECR. 
The inclusion of a local galactic source for the UHECR origin is 
further supported by the experimental evidences of an UHECR excess 
around $10^{18}$ eV from the direction of the galactic center region. 
This excess can be considered as the peak of a gaussian distribution 
of the second component and permits us to make a calibration among the 
GRB flux and the UHECR energies.

Perhaps the main constraint to the TCM is, if the UHECR sources are 
close, why the arrival direction of the events does not point toward 
their sources? So far, the UHECR distribution does not follow the 
galactic star distribution. This means that probably its surrounding 
diffuse halo and even the local distribution of galaxies can lodge 
UHECR sources. UHECR event rates beyond the GZK cutoff as observed by 
AGASA experiment can be linked at least in part to GRBs of short-hard 
type, because these GRBs neither have afterglows nor redshift suggesting 
that some of these GRBs sources are close to or inside of our galaxy. 
These results are in agreement with others surveys \cite{cline05}, showing   
that GRBs with time duration below $100$ ms appear to form a separate class 
of GRBs, because from their asymmetry plot the events appear to originate 
nearby within the Galaxy.

The short-hard GRBs probably are formed for instance by coalescence of 
corotating binary neutron star systems. This process is dominated by a 
strong magnetic field. Thus, magnetic deflections of the UHECR in the 
first stages of their propagation can be responsible of a arrival 
direction of events without pointing toward their sources.

We are waiting for the next round of the AUGER experiment. Certainly 
only after a large set of data that we can confirm or refute the 
TCM for the origin of UHECR.

\section{Acknowledments}

This work was partially supported by FAPERJ (The Research Fostering 
Foundation of the State of Rio de Janeiro) and CNPq (The National 
Council of Research and Development) of Brazil. 


\begin{figure}[th]
\includegraphics[clip,width=1.0
\textwidth,height=1.0\textheight,angle=0.] {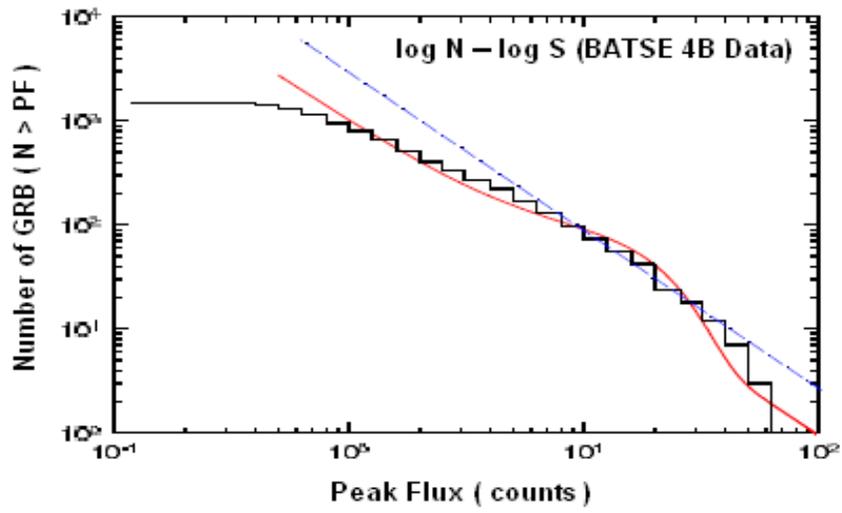}
\caption{The $\log N \rightarrow \log S(E)$ plot. Black line: GRBs distribution 
on the basis of 2704 burst from BATSE catalog. Blue line: Uniform distribution. 
Red line: Uniform distribution plus gaussian distribution (Two component model)}%
\end{figure}

\begin{figure}[th]
\includegraphics[clip,width=0.9
\textwidth,height=0.9\textheight,angle=0.] {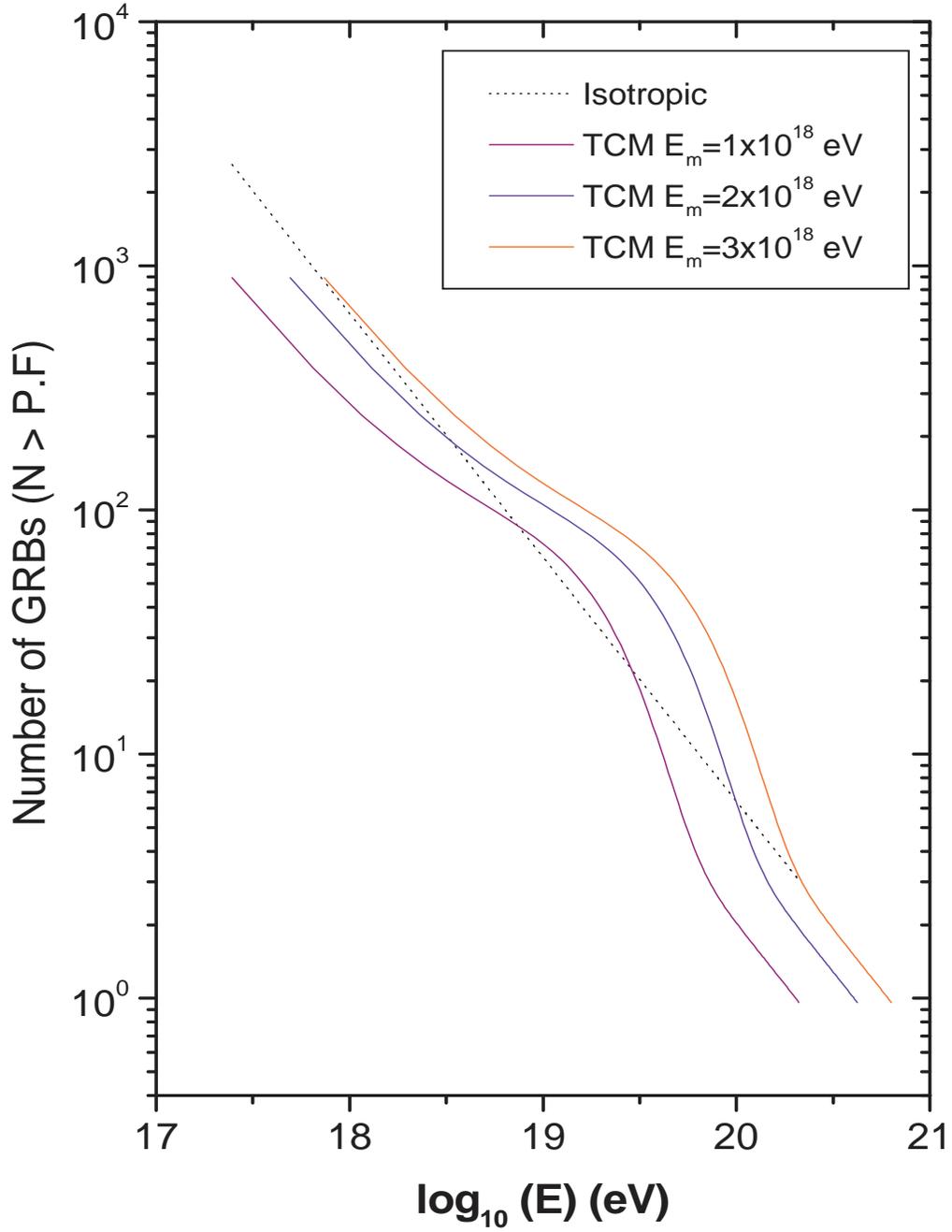}
\caption{Two component model cumulative distribution for UHECR (one by one 
association with GRBs) and for three different energies in the peak of the 
gaussian distribution }%
\end{figure}

\begin{figure}[th]
\includegraphics[clip,width=0.9
\textwidth,height=0.9\textheight,angle=0.] {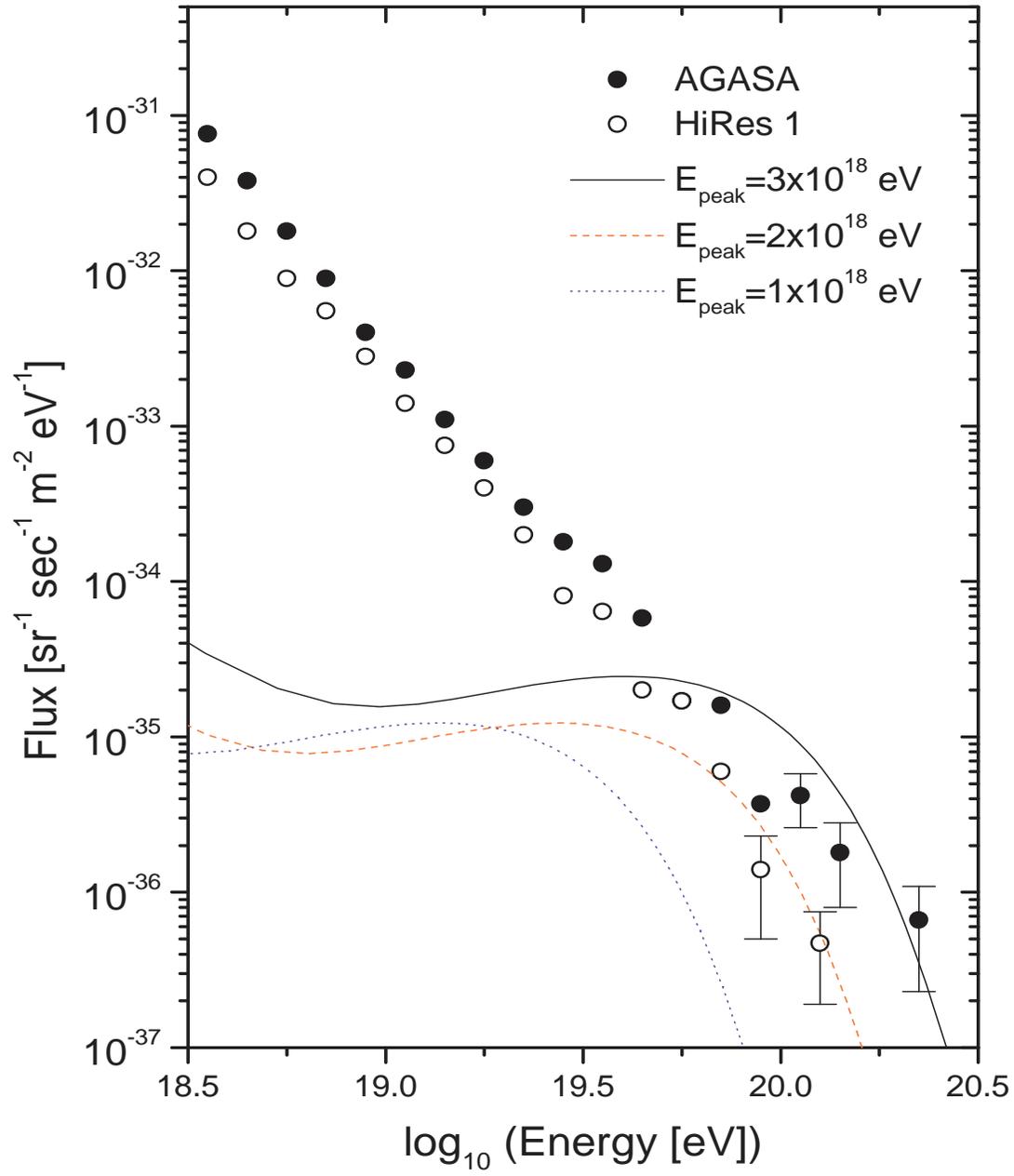}
\caption{Comparison between the Two Component Model and experimental data}%
\end{figure}

\end{document}